# Panel and Pseudo-Panel Estimation of Cross-Sectional and Time Series Elasticities of Food Consumption:

# The Case of American and Polish Data

François Gardes
Université de Paris I Panthéon-Sorbonne (Cermsem)

Greg J. Duncan
Northwestern University

Patrice Gaubert
Université du Littoral (LEMMA)

Marc Gurgand
CNRS – Delta

Christophe Starzec
CNRS – TEAM

The problem addressed in this article is the bias to income and expenditure elasticities estimated on pseudo-panel data caused by measurement error and unobserved heterogeneity. We gauge empirically these biases by comparing cross-sectional, pseudo-panel and true panel data from both Polish and American expenditure surveys. Our results suggest that unobserved heterogeneity imparts a downward bias to cross-section estimates of income elasticities of at-home food expenditures and an upward bias to estimates of income elasticities of away-from-home food expenditures. "Within" and first-difference estimators suffer less bias, but only if the effects of measurement error are accounted for with instrumental variables.

# 1. INTRODUCTION

A multitude of data types and econometric models can be used to estimate demand systems. Data types include aggregate time series, within-group time series, cross-sections, pseudo-panels using aggregated data, and cross sections and panels using individual data. Aggregate time series data frequently produce aggregation biases because of composition effects due to the change of the population or the heterogeneity of price and income effect between different social classes. These problems have led the vast majority of empirical studies in labor economics to use individual data (Angrist and Krueger, 1998).

On the other hand, *individual* panel data generally span short time periods and are subject to nonresponse attrition bias. Even panels on countries or industrial sectors can suffer from structural changes or composition effects that make it difficult to maintain the stationarity hypotheses for all variables.

Thus, grouping data to estimate on pseudo-panel is an alternative, even when panel data exist, in order to estimate on longer periods or to compare different countries. Pseudo-panel data are typically constructed from a time series of independent surveys which have been conducted under the same methodology on the same reference population, but in different periods, sometimes consecutive and sometimes not.

In pseudo-panel analyses, individuals are grouped according to criteria that do not change from one survey to another, such as their birth year or the education level of the reference person of a household. Estimation with pseudo-panel data diminishes efficiency on the cross-section dimension, but we will show that it also gives rise to a heteroscedasticity in the time dimension.

Static and dynamic demand models have been developed for these different types of data, with each adopting a different approach to problems caused by unobserved heterogeneity across consumption units or time period of measurement as well as the cross-equation restrictions imposed by consumption theory. The use of different types of data helps reveal the nature of the biases they impart to estimates of income and expenditure elasticities.

This article addresses the issue of bias to income and expenditure elasticities caused by errors of specification, measurement and omitted variables, and by heteroscedasticity, in grouped and individual-based models. We gauge these biases by estimating static expenditure models using cross-sectional, pseudo-panel and true panel data from both Polish and U.S. expenditure surveys. It is, to our knowledge, the first comparison between cross-sectional, pseudo-panel and panel estimations based on the same data set. The use of one of our two data sets (the Panel Study of Income Dynamics - PSID) is motivated by the numerous expenditure studies based on it (Altug and Miller, 1990; Altonji and Siow, 1987; Hall and Mishkin, 1982; Naik and Moore, 1996; Zeldes, 1989). Our second data set is from Poland in the late 1980s, which enables us to capitalize on large income and price variations during the transition period in Poland.

Section 2 presents a background discussion. The econometric problems and methods used are presented in Section 3. The data are described in the fourth section, with results presented in the fifth section and discussed in the sixth section.

## 2. BACKGROUND

No matter how complete, survey data on household expenditures and demographic characteristics lack explicit measures of all of the possible factors that might bias the estimates of income and price elasticities. For example, the value of time differs across households and is positively related to a household's observed income. Since consumption activities (e.g., eating meals) often involve inputs of both goods (e.g., groceries) and time (e.g., spent cooking and eating), households will face different (full) prices of consumption even if the prices of the goods-based inputs are identical. If, as is likely in the case of meals prepared at home, these prices are positively associated with income and themselves have a negative effect on consumption, the omission of explicit measurement of full prices will impart a negative bias to the estimated income elasticities. The same argument can be applied to the case of virtual prices arising, say, from liquidity constraints that are most likely in low-income households (Cardoso and Gardes, 1997). Taking into account the virtual prices appearing from non-monetary resources such as time, or restriction of the choice space due to constraints applying only to sub-populations, or are changing from one period to another, may help to better identify and understand cross-sectional and time series estimation differences.

Panel data on households provide opportunities to reduce these biases, since they contain information on changes in expenditures and income for the same households. Differencing successive panel waves nets out the biasing effects of unmeasured persistent characteristics. But while reducing bias due to omitted variables, differencing income data is likely to magnify another source of bias: measurement error. Altonji and Siow (1987) demonstrate the likely importance of measurement error in the context of first-difference consumption models by showing that estimates of income elasticities are several times higher when income change is instrumented than when it is not.

Deaton (1986) presents the case for using "pseudo-panel" data to estimate demand systems. He assumes that the researcher has independent cross sections with the required expenditure and demographic information and shows how cross sections in successive years can be grouped into comparable demographic categories and then differenced to produce many of the advantages gained from differencing individual panel data. Grouping into cells tends to homogenize the individual effects among individuals grouped in the same cell, so that the average specific effect is approximately invariant between two periods, and it is efficiently removed by within or first-differences transformations.

We evaluate implications of alternative approaches to estimating demand systems using two sets of household panel expenditure data. The two panels provide us with data needed to estimate static expenditure models in first difference and "within" form. However, these data can also be treated as though they came from independent cross sections and from grouped rather than individual-household-level observations. Thus we are able to compare estimates from a wide variety of data types. Habit persistence and other dynamic factors give rise to dynamic models (e.g., Naik and Moore, 1996). We estimated dynamic versions of the static models using usual instrumentation methods (Arrelano, 1989) and found that elasticity estimates were quite similar to those estimated for the static models that are presented in this paper. However these dynamic versions are questionable as far as the specification and the econometric problems are concerned. So we prefer to consider only the static specification.

True panel and pseudo-panel methods each offers advantages and disadvantages for handling the estimation problems inherent in expenditure models. A first set of concerns centers on measurement error. Survey reports of household income are measured with error; differencing reports of household income across waves undoubtedly increases the extent of error. Instrumental variables can be used to address the biases caused by measurement error (Altonji and Siow, 1987). Like instrumentation, aggregation in pseudo-panel data helps to reduce the biasing effects of measurement error, so we expect that the income elasticity parameters estimated with pseudo-panel data will be similar to those estimated on instrumented income using true panel data. Since measurement error is not likely to be serious in the case of variables like location, age, social category, and family composition, we confine our instrumental variables adjustments to our income and total expenditure predictors. Measurement errors in our dependent expenditure variable are included in model residuals and, unless correlated with the levels of our independent variables, should not bias the coefficient estimates.

Special errors in measurement can appear in pseudo-panel data when corresponding cells do not contain the same individuals in two different periods. Thus, if the first observation for cell 1 during the first period is an individual A, it will be paired with a similar individual B observed during the second period, so that measurement error arises between this observation of B and the true values for A if he or she had been observed during the second period.

Deaton (1986) treats this problem as a measurement error: sample-based cohort averages are error-ridden measures of true cohort averages. He proposes a Fuller-type correction to ensure convergence of pseudo-panel estimates. However, Verbeek and Nijman (1993) show that Deaton’s estimator converges with the number of time periods. Moreover, Lepellec and Roux (2002) show that the measurement error correction variance matrix used in Deaton’s estimator is often not definite positive in the data. The simpler pseudo-panel estimator used in this paper has been shown to converge with the cell sizes (Verbeek and Nijman, 1993, Moffit, 1993) because measurement error becomes negligible when cells are large. Based on simulations, Verbeek and Nijman (1993) argue that cells must contain about one hundred individuals, although the cell sizes may be smaller if the individuals grouped in each cell are sufficiently homogeneous.

Resolving the measurement error problem by using large samples within cells creates another problem – the loss of efficiency of the estimators. This difficulty was shown by Cramer (1964) and Haitovsky (1973) with estimations based on grouped data and by Pakes (1983) with the problem induced by an omitted variable with a group structure, which is similar to the problem of measurement error.

The answer to the efficiency problem is to define groupings that are optimal in the sense of keeping efficiency losses to a minimum but also keeping measurement error ignorably small (Baltagi, 1995). Grouping methods were developed by Cramer, Haitovsky and Theil (1967), and again in Verbeek and Nijman (1993) and involve the careful choice of cohorts in order to obtain the largest reduction of heterogeneity within each cohort but at the same time maximizing the heterogeneity between them. Following these empirical principles, the use of pseudo-panels leads to consistent and efficient estimators without the problems associated with true panels. Our own work below groups individuals into cells that are both homogeneous and large.

Second, the aggregation inherent in pseudo-panel data produces a systematic heteroscedasticity. This can be corrected exactly by decomposing the data into between and within dimensions and computing the exact heteroscedasticity on both dimensions. But since the heteroscedastic factor depends on time, correcting it by GLS makes individual specific effects vary with time, thus canceling the spectral decomposition in between and within dimensions. This can result in serious estimation errors (Gurgand, Gardes and Bolduc, 1997).

The approximate correction of heteroscedasticity that we use consists in weighting each observation by a heteroscedasticity factor that is a function of, but not exactly equal to, cell size. Thus the LS coefficients computed on the grouped data may differ slightly from those estimated on individual data. As described in the next section, this approximate and easily implemented correction uses GLS on the within and between dimensions with a common variance-covariance matrix computed as the between transformation of the heteroscedastic structure due to aggregation.

Third, unmeasured heterogeneity is likely to be present in both panel and pseudo-panel data. In the case of panel data, the individual-specific effect for household h is $\alpha(h)$, which is assumed to be constant through time. In the case of pseudo-panel data, the individual-specific effects for a household (h) belonging to the cell (H) at period *t*, can be written as the sum of two orthogonal effects: $\alpha(h,t)=\mu(H) + \upsilon(h,t)$. Note that the second component depends on time since the individuals composing the cell H change through time.

The specific effect $\mu$ corresponding to the cell H ($\mu(H)$) represents the influence of unknown explanatory variables W(H), constant through time, for the reference group H, which is defined here by the cell selection criteria. $\upsilon(h,t)$ are individual specific effects containing effects of unknown explanatory variables Z(h,t). In the pseudo-panel data the aggregated specific effect $\zeta(H)$ for the cell H is defined as the aggregation of individual specific effects:

$$\zeta(H,t)=\sum \gamma(h,t)*\alpha(h,t) = \mu(H) + \sum\gamma(h,t) * \upsilon(h,t)$$

where t indicates the observation period and $\gamma$ is the weight for the aggregation of h within cells. Note that the aggregate but not individual specific effects depend on time.

The within and first difference operators estimated with panel data cancel the individual specific effects $\alpha(h)$. The component $\mu(H)$ is also canceled on pseudo-panel data by the same operators, while the individual effect $\upsilon(h,t)$ may be largely eliminated by the aggregation. Thus it can be supposed that the endogeneity of the specific effect is greater on individual than on aggregated data, as aggregation cancels a part of this effect.

Therefore, with panel data the within and the first-differences operators suppress all the endogeneity biases. With pseudo-panel data the same operator suppresses the endogeneity due to $\mu$, but not that due to $\sum\gamma(h,t) * \upsilon(h,t)$. For each individual this part of the residual may be smaller relatively to $\mu$, as cell homogeneity is increased. Conversely, the aggregation into cells is likely to cancel this same component $\upsilon$ across individuals, so that it is not easy to predict the effect of the aggregation on the endogeneity bias.

Our search for robust results is facilitated by the fact that the two panel data sets we use cover extremely different societies and historical periods. One is from the United States for 1984-1987, a period of steady and substantial macroeconomic growth. The second source is from Poland for 1987-1990, a turbulent period that spans the beginning of Poland’s transition from a command to free-market economy.

## 3. SPECIFICATION AND ECONOMETRICS OF THE CONSUMPTION MODEL

Data constraints force us to estimate a demand system on only two commodity groups over a period of four years: food consumed at home and food consumed away from home. In addition, away-from-home food expenditures are rare in the Polish data so that our estimates are not very reliable, but we keep them in order to compare them with PSID estimates. We use the Almost Ideal Demand system developed by Deaton and Muellbauer (1980), with a quadratic form for the natural logarithm of total income or expenditures in order to take into account nonlinearities.

Note that the true quadratic system proposed by Banks et al. (1997) implies much more sophisticated econometrics if the non-linear effect of prices is taken into account. It may be difficult to estimate precisely the price effect because of the short duration estimation period in the PSID data. On the other hand, our Polish data contains relative prices for food which change both across sixteen quarters and between four socio-economic classes. Thus we estimated the linearized version of QAIDS (with the Stone index) on the Polish panel using the convergence algorithm proposed by Banks et al. (1997) to estimate the integrability parameter e(p) in the coefficient of the quadratic log income (see equation 1). We obtained very similar income elasticities for food at home and food away to those obtained by linear AI Demand System. We present only these AI estimates for both countries in table I and II. The additivity constraint is automatically imposed by OLS.

The possible correlation between the residuals of food at home and food away would suggest the use of Seemingly Unrelated Regression. We tested this possible correlation on Polish data and found no significant difference between OLS and SUR estimations. Since relative prices do not vary much within waves of the same survey compared to the variations between different years, at least for the U.S. in the mid 1980s, even if we consider quarterly variations of prices, we account for price effects and other macro-economic shocks with survey year dummies for the US data. For the Polish data each individual was given a price index differentiated by the social category and the quarter of the year in which he was surveyed.

Our model takes the following form:

$$w_{ht}^{i} = a^{i} + b^{i} \ln\left(Y_{ht} / p_{t}\right) + c^{i} / e\left(p\right)\left[\ln\left(Y_{ht} / p_{t}\right)\right]^{2} + Z_{ht} d^{i} + u_{ht}^{i} \qquad (1)$$

with $w^{i}_{ht}$ the expenditure budget share on good *i* by household *h* at time t, $Y_{ht}$ its income (in the case of one of our U.S. data or logarithmic total expenditure in the case of the other — the Polish expenditure panel), $p_t$ the Stone price index, $Z_{ht}$ a matrix of socio-economic characteristics and survey year or quarter dummies and

$$e(p) = \prod_{i} p_{it}^{b_i}$$

is a factor estimated by the convergence procedure proposed by Banks et al. (1997) which ensures the integrability of the demand system. When using total expenditure data from the Polish panel, the allocation of income between consumption and saving can be ignored, and total expenditures can be considered as a proxy for permanent income. Our U.S. data do not provide information on total expenditure, so that income elasticities are computed on the basis of total household disposable income (the use of income instead of total expenditures would be better served by a model in which income is decomposed into permanent and transitory components).

Our cross-sectional estimates of equation (1) are based on data on individual households from each available single-year cross-section (1984-1987 in the case of the PSID and 1987-1990 in the case of the Polish expenditure survey).

First difference and within operators are common procedures employed to eliminate biases caused by persistent omitted variables, and we use our panel data to obtain first-difference and within estimates of our model. Following Altonji and Siow (1987), we estimate our models both with and without instrumenting for change in log income or expenditures.

Instrumenting income from the PSID is necessary because of likely measurement errors observed in such income data. We also instrument the total expenditure from the Polish surveys because measurement errors for both total expenditures and food expenditures are likely correlated.

In the QAIDS specification the classical errors-in-variables cannot hold for the squared term if it holds for the log of income. As far as we know this problem has not been yet solved conveniently so we simply used the square of the instrumented income, checking that a separate instrumentation of the squared term does not change significantly the results.

For cross-sections and first-differences we found two types of correlations: between individuals in cross-sections and between periods in first-differences. We consider this problem by estimating separately for each period with a robust OLS method. For the within estimation, all autoregressive processes on the residuals (for instance resulting from partial adjustment in exogenous variables) are taken into account, as suggested by Hsiao (1986, p.95-96), by estimating the system of equations written for the successive periods**.**

Pseudo-panel estimates. The grouping of data for pseudo-panels is based on six age cohorts and two or three education levels. The grouping of households (h,t) in the cells (H,t) gives rise to the exact aggregated model:

$$\sum_{h\in H}\gamma_{ht}w_{ht}^{i}=w_{Ht}^{i}=\left(\sum_{h}\gamma_{ht}X_{ht}\right)A^{i}+\alpha_{H}^{i}+\sum_{h}\gamma_{ht}\varepsilon_{ht}^{i}$$

with $\gamma_{ht}=\frac{Y_{ht}}{\sum_{h\in H}Y_{ht}}$ under the hypothesis $\alpha_{h}^{i}=\alpha_{H}^{i}$ for $h\in H$ (a natural hypothesis, according to the grouping of households into a same H cell). A heteroscedasticity factor $\delta_{Ht}=\sum_{h\in H}\gamma_{ht}^{2}$ arises for the residual $\varepsilon^i$, which is due to the change of cells sizes (as $\gamma \cong \frac{1}{|H|}$ if the two grouping criteria homogenize the household's total expenditures). Thus, the grouping of data builds up a heteroscedasticity which may change through time, because of the variation of the cells sizes. We show in Appendix A that this heteroscedasticity cannot be corrected by usual methods.

We present exact correction procedures in Appendix A and show that, under a symmetry condition, heteroscedasticity can also be approximately corrected by simple generalized least squares based on the average heteroscedasticity factor over time for each cell:

$$\sum_{t=1}^{T}\sum_{h\in H}\frac{1}{T}\gamma_{ht}^{2} = \delta_H$$

In our data sets, the size variation through time for each cell is unimportant, so that the heteroscedasticity factor due to the grouping is quite invariant. In this paper, heteroscedasticity is corrected by the exact procedure for within and between estimations, and also by simple generalized least squares based on the average heteroscedasticity factor for all estimations.

For PSID data, the population is randomly divided into four sub-samples, each of which is used to aggregate data for the different years. This prevents the same household from being included in the same cell in more than one period (in which case the aggregation would just correspond to grouped panel data).

For Polish data, all households (after filtering for some outliers defined on cross-section estimations) in the cross-sectional component of each survey are used for the pseudo-panelization; panel households belonging to the surveys are excluded. Sample sizes for each year are around 27,000 households, which is much larger than for the PSID data.

The PSID cells sizes vary from 9 to 183 households with a mean of 65.5 and from 8 to 60, with a mean of 25.1 for the Polish data. Fourteen of the 72 cells constituting the whole pseudo-panel in the PSID contain less than 30 households, representing only 4 % of the whole population. As the correction for the heteroscedasticity on the pseudo-panel data consists in weighting each cell by weights close to its size, the estimation without these small cells gives the same results than those for the whole. For each cell the size variation through time is much less important, so that the heteroscedasticity factor due to the grouping is quite invariant through time.

It is clear that the residuals for two adjacent equations estimated in first differences, ($u^i_{h,t}$-$u^i_{h,t-1}$) and ($u^i_{h,t-1}$-$u^i_{h,t-2}$), are systematically correlated. Since all specifications are estimated by Zellner's seemingly unrelated regressions, our procedures take into account the correlation between the residuals of the two food components.

Price effects are taken into account by period dummies for the PSID and by price elasticities for Poland. The age of the household's head, and family size and structure are also taken into account in the estimations. Adding other control variables such as head's sex, education level, wealth, and employment status in the PSID had very little effect on the estimates. We selected only age and family structure variables for the PSID to make the estimations comparable to the results based on the Polish data.

Correction for grouped heteroscedasticity may still leave some heteroscedasticity for the estimations at the individual level. We test for this by regressing the squared residuals on a quadratic form of explanatory variables, thus correcting it when necessary by weighting all observations by the inverse absolute residual. The coefficient on the squared income is generally significant, but QAIDS estimates are very close to AIDS.

## 4. DATA

The Panel Study of Income Dynamics. Since 1968, the PSID has followed and interviewed annually a national sample that began with about 5,000 U.S. families (Hill, 1992). The original sample consisted of two sub-samples: i) an equal-probability sample of about 3,000 households drawn from the Survey Research Center's dwelling-based sampling frame; and ii) a sample of low-income families that had been interviewed in 1966 as part of the U.S. Census Bureau's Survey of Economic Opportunity and who consented to participate in the PSID.

When weighted, the combined sample is designed to be continuously representative of the nonimmigrant population as a whole. To avoid problems that might be associated with the low-income sub-sample, our estimations based on individual-household data are limited to the (unweighted) equal-probability portion of the PSID sample. To maximize within-cell sample sizes, our pseudo-panel estimates are based on the combined, total weighted PSID sample. We note instances when pseudo-panel estimates differed from those based on the equal-probability portion of the PSID sample.

Since income instrumentation requires lagged measures from two previous years, our 1982-87 subset of PSID data provides us with data spanning five cross sections (1983-1987). We use only four years in the estimation of the consumption equation to be comparable with the Polish data. In all cases the data are restricted to households in which the head did not change over the six-year period and to households with major imputations on neither food expenditure nor income variables (in terms of the PSID's "Accuracy" imputation flags, we excluded cases with codes of 2 for income measures and 1 or 2 for food at home and food away from home measures).

In order to construct cohorts for the pseudo-panels, we defined a series of variables based on the age and education levels of the household head. Specifically, we define : i) 6 cohorts of age of household head: under 30 years old, 30-39, 40-49, 50-59, 60-69, and over 69 years old; and ii) three levels of education of household head: did not complete high school (12 grades), completed high school but no additional academic training, and completed at least some university-level schooling.

The PSID provides information on two categories of expenditure: food consumed at home and food consumed away from home and has been used in many expenditure studies (e.g., Hall and Mishkin, 1982; Altonji and Siow, 1987; Zeldes, 1989; Altug and Miller, 1990; Naik and Moore, 1996). These expenditures are reported by the households as an estimation of their yearly consumption so reporting zero consumption can be considered as a true no-consumption. That is why no correction of selection bias is needed.

All of these studies were based on the cross-section analyses and thus may be biased because of the endogeneity problems discussed above. To adjust expenditures and income for family size we use the Oxford equivalence scale: 1.0 for the first adult, 0.8 for the others adults, 0.5 for the children over 5 years old and 0.4 for those under 6 years old. Our expenditure equations also include a number of household structure variables to provide additional adjustments for possible expenditure differences across different family types.

Disposable income is computed as total annual household cash income plus food stamps minus household payments of alimony and child support to dependents living outside the household and minus income taxes paid (the household's expenditure on food bought with food stamps is also included in our measure of at-home food expenditure). As instruments for levels of disposable income we follow Altonji and Siow (1987) in including three lags of quits, layoffs, promotions and wage-rate changes for the household head (as with Altonji and Siow (1987), we construct our wage rate measure from a question sequence about rate of hourly pay or salary that is independent of the question sequence that provides the data on disposable household income) as well as changes in family composition other than the head, marriage and divorce/widowhood for the head, city size and region dummies. For first-difference models, the change in disposable income is instrumented using the first-difference of instrumented income in level.

Means and standard deviations of the PSID variables are presented in Appendix Table 1; coefficients and standard errors from the first stage of the instrumental variables procedure are presented in Appendix Table 2.

The Polish expenditure panel. Household budget surveys have been conducted in Poland for many years. In the analyzed period (1987-1990) the annual total sample size was about 30 thousand households; this is approximately 0.3% of all the households in Poland. The data were collected by a rotation method on a quarterly basis. The master sample consists of households and persons living in randomly selected dwellings. To generate it, a two stage, and in the second stage, two phase sampling procedure was used. The full description of the master sample generating procedure is given by Lednicki (1982).

Master samples for each year contains data from four different sub-samples. Two sub-samples began their interviews in 1986 and ended the four-year survey period in 1989. They were replaced by new sub-samples in 1990. Another two sub-samples of the same size were started in 1987 and followed through 1990.

Over this four-year period it is possible to identify households participating in the surveys during all four years (these households form a four-year panel. There is no formal identification possibility (by number) of this repetitive participation, but special procedures allowed us to specify the four year participants with a very high probability. The checked and tested number of households is about 3,707 (3,630 after some filtering). The available information is as detailed as for the cross-sectional surveys: all typical socio-demographic characteristics of households and individuals, as well as details on incomes and expenditures, are measured. The expenditures are reported for three consecutive months each year, so we considered again that zero expenditure is a true no-consumption case. So no correction is needed for selection bias, like for the PSID**.**

Comparisons between reported household income and record-based information showed a number of large discrepancies. For employees of state-owned and cooperative enterprises (who constituted more than 90% of wage-earners until 1991), wage and salary incomes were checked at the source (employers). In a study by Kordos and Kubiczek (1991), it was estimated that employees' income declarations for 1991 were 21% lower, on average, than employers' declaration. Generally, the proportion of unreported income is decreasing with the level of education and increasing with age. In cases where declared income was lower than that reported by enterprises, household's income was increased to the level of the reported one. Since income measures are used only to form instrumental variables in our expenditure equations the measurement error is likely to cause only minor problems.

Appendix Table 3 presents descriptive information on the Polish data, while Appendix Table 4 presents coefficients from the instrumental-variables equation. The period 1987-1990 covered by the Polish data is unusual even in Polish economic history. It represents the shift from the centrally planned, rationed economy (1987) to a relatively unconstrained fully liberal market economy (1990). GDP grew by 4.1% between 1987 and 1988, but fell by .2% between 1988 and 1989 and by 11.6% between 1989 and 1990. Price increases across these pairs of years were 60.2%, 251.1% and 585.7%, respectively. Thus, the transition years 1988 and 1989 produced a period of a very high inflation and a mixture of free-market, shadow and administrated economy.

This means that the consumers' market reactions could have been highly influenced by these unusual situations. This is most likely the case of the year 1989 when uncertainty, inflation, market disequilibrium and political instability reached their highest level. Moreover, in 1989 and 1990 individuals were facing large real income fluctuations as well as dramatic changes in relative prices. This unstable situation produced atypical consumption behaviors of households facing a subsistence constraint. This may be the case of very low income households having faced a dramatic decrease of their purchasing power (over 30%).

## 5. RESULTS

Estimates from our various models are presented in Tables 1 (PSID) and 2 (Polish surveys). Respective columns show income (for PSID; total expenditure for Polish data) elasticity estimates for between, cross section (computed as the means of cross-sectional estimates obtained on each cross-sectional survey), within and first-difference models. Results are also presented separately for models in which income (total expenditure) is and is not instrumented using the models detailed in Appendix Tables 2 and 4. We expect the between estimates to be similar to the average of cross-section estimates. Compared to the within estimates, the first-difference estimates may be biased by greater measurement error, but the specific effects may be better taken into account whenever they change within the period.

For the PSID we eliminated some observations to obtain robust estimations using the DFBETAS explained by Belsley , Kuh and Welsch (1980) to select outliers. We eliminated observation when $\text{DFBETAS} > 2/\sqrt{n}$, where n is the number of observations. Rejected observations represent 4% of the sample.

For Polish data the estimation of a Quadratic AI demand system by iteration on the integrability parameter (see Banks et al., 1999) gives very similar results, except for Food away (between and within estimators); QAIDS estimations are very close to the results presented in Table 2. Filtering data for outliers (like for PSID) did not change significantly the results.

For pseudo-panel data heteroscedasticity has been corrected by the approximate method (GLS with a heteroscedasticity factor $\delta_H$, which is constant through time, Table 1 and Table 2, a), and the exact method (Table 1 and Table 2, b) presented in Appendix A. We present also for the Polish pseudo-panel the estimates obtained without correction (Table 2, c) and with a false correction (GLS with a heteroscedasticity factor $\delta_{Ht}$ ,Table 2, d). The between and cross-section estimates are similar for the different correction methods, especially for food at home, but the within and first-differences estimates obtained under the false correction, which is currently used in pseudo-panel estimations, gives very different estimates than those computed for the approximate correction, the exact one or no correction. So, the correction for heteroscedasticity seems to be an important methodological point to address in the estimations on pseudo-panel data. The false correction gives very different estimates, especially for time series estimations. However in our case, the exact correction gives rise to estimated parameters which are close to those obtained by the approximate correction, so we discuss principally these estimates that can be easily compared under the spectral decomposition into the between and within dimensions.

Looking first at the PSID results for at-home food expenditures, it is quite apparent that elasticity estimates are very sensitive to adjustments for measurement error and unmeasured heterogeneity. Cross-sectional estimates of at-home income elasticities are low (between .15 and .30) but statistically significant without or with instrumentation (when performing robust estimations). The between estimates effectively average the cross sections and also produce low estimates of elasticities. Pseudo-panel data produces similar elasticities for between and cross-sections estimates. Despite some variations between the different estimations, the relative income elasticity of food at home is around .20 based on this collection of methods.

Within and first difference estimates of PSID-based income elasticities are around 0 without instrumentation and around .40 with instrumentation. A Hausman test strongly rejects (p-value<.01) the equality of within and between estimates (Table 3). The test compares within and GLS estimates: equivalently it can be built from the within and between estimates (see Baltagi, 1995, p. 69). The test is computed by the usual quadratic form, distributed as a $\chi^2$, with V defined as the variance of the difference between the estimators tested: $(\beta_b - \beta_w)'$ $(V^{-1})$ $(\beta_b - \beta_w)$ where $\beta = (\beta_{ly})$ or $\beta = (\beta_{ly}, \beta_{ly^2})$ for the quadratic estimation on the Polish panel, $V=V_b+V_w$ corresponds to all the explanatory variables. Note that a test with V as a matrix 2x2 computed only for the two income variables would be biased.

Since the within and first difference models adjust for persistent heterogeneity and the instrumentation adjusts for measurement error, .40 is our preferred approximate estimate of the income elasticity of at-home food expenditures in the United States. This is roughly double the size of the corresponding between and cross-sectional parameter estimates, which suggests that failure to adjust for heterogeneity imparts a considerably downward bias to cross-sectional estimates. PSID-based pseudo-panel data also produce a significantly (according to a Hausman test) higher elasticity estimates for first-differences as compared with between and cross-sections models although there is little consistency across the full set of pseudo-panel estimates.

Expenditure elasticities for at-home food estimated with the Polish data are much higher in value than the income elasticity estimates based on PSID data (note that the Polish elasticities are computed on total expenditures so that they must be multiplied by the income elasticity of total expenditures, which is around .7, to be compared to the PSID income elasticities). Higher elasticities are to be expected for a country in which food constitutes a share of total expenditures that is three times higher than in the U.S. (Appendix Tables 1 and 3). Their consistency probably stems from the smaller degree of measurement error in the Polish expenditure as opposed to the PSID income data. Before grouping into cells ten households have been suppressed. The criterion was the prediction of food consumption in cross-section estimations. On the panel data, robust estimations produced by suppressing some outliers gave similar results as those obtained for the whole panel. Time series Polish pseudo-panel estimates are a little smaller than estimates based on the microdata. On the whole the estimations on Polish data also produce higher within and first-differences elasticities than between and cross-sections.

PSID-based income elasticities for away-from-home food expenditures are quite different from and even more sensitive to specification than the at-home elasticity estimates. Between and cross-sectional estimates are around 1.0 in both individual and pseudo-panel data. In contrast to the case of at-home expenditures, adjustments for heterogeneity through use of within and first difference estimates produce much lower estimates. We speculate on why this might be the case in our discussion section.

Polish food-away expenditures are relatively rarely reported in the survey and they are very low even when compared with other countries with comparable income levels. Moreover, about 70% of this expenditure in the observed periods is spent on highly subsidized business canteens and cafeterias. So the estimations should be compared with caution with the food away expenditure estimation in other countries.

Price data in Poland enabled us to compute price elasticities. Quarterly price indices for four social categories were computed from monthly GUS (Polish Main Statistical Office) publication Biuletyn Statystyczny and imputed at the individual level to the data set. The variability of the prices both over time and over social categories provides good estimates of the direct compensated elasticity for food at home (around minus one). Contrary to the income elasticities, the cross-section estimates of direct price elasticities are close to those obtained from time-series. As prices change between quarters and households of different types the only explanation of an endogeneity bias would be a correlation between household types and level of prices. Such a correlation is less probable than the one between the relative income and the specific component of food consumption (which produces the endogeneity bias on income elasticities). However, if systematically high (or low) price level is correlated with household's type (which can be the case for instance on segmented markets) and if these types of households are characterized by a systematically positive or negative specific consumption, the endogeneity bias can appear. This is not the case of Poland in 1987-1990.

## 6. DISCUSSION

We have attempted to assess the bias to income and expenditure elasticity estimates caused by inattention to measurement error and unobserved heterogeneity. In the case of the U.S. at-home food expenditure elasticity, our preferred estimate is around .40. Failure to adjust for unmeasured heterogeneity and, in some cases, measurement error appears to impart a substantial downward bias to this estimate. These adjustments operate in the same direction for estimates of the at-home expenditure elasticity found in the Polish data.

In the case of U.S. away-from-home food expenditures, our preferred elasticity estimate is less certain but similar in magnitude to the .40 elasticity for at-home expenditures. (Note that it is considerably higher in pseudo-panel estimate). Surprising here is the magnitude and sign (upward) of the apparent bias inherent in both individual and pseudo-panel estimates that do not adjust for unobserved heterogeneity.

Why should unmeasured heterogeneity induce an upward bias in away-from-home expenditures in the U.S. and Polish data? Earlier, we speculated that a likely downward bias for the at-home food elasticity estimates may be caused by failure to account for the fact that the value of time differs across households and is positively related to the household's observed income. The time input to producing at-home meals lead households to face different (full) prices of consumption even if the prices of the goods-based inputs are identical. If these prices are positively associated with income and themselves have a negative effect on consumption, their omission will impart a negative bias to the estimated income elasticities. Note that the bias seems somewhat less pronounced on pseudo-panel data. The correlation between income and the specific effect may be decreased by aggregation, as suggested in section 2.

In the case of expenditures for meals consumed in restaurants, there are large variations in the mixture of food and service components. Time spent consuming full-service restaurant meals is typically longer than time spent consuming fast-food restaurant meals, but in this case higher-income households may well attach a more positive value to such time than low income households. Failing to control for this source of heterogeneity will probably impart a positive bias to estimates of income elasticities (note that the endogeneity bias is much more pronounced for food away compared to food at home).

The use of time can be measured with the shadow price of time. This shadow price increases with household income, so that the complete price for food, computed as the sum of the monetary and shadow prices, increases also along the income distribution. Therefore, the difference between cross-section and time-series income elasticities can be related to the change of this complete price. The argument is formalized in Appendix B.

Calibrating the price effect as half of the income effect (as suggested by Frisch) and using equation (B.2) in Appendix B produces an estimate of the income elasticity of the food shadow price (Table 4). It is remarkable that these elasticities are similar in the two countries, positive and around 1 for food at home, and negative and much larger for food away. Thus, the time constraint imparts much stronger relative changes in expenditures in restaurants than at home, which is normal since the budget shares of food away are much smaller in both countries (specially in Poland, where the ratio of the price elasticities is the largest): a substitution between food at home and food away when the ratio of their complete prices change imparts a similar change in the expenditures, but greater changes in the food away budget share.

Elasticity estimates from the Polish data also proved somewhat sensitive to adjustments for heterogeneity. This is not surprising given the very different price regimes in the Polish economy during this period, with one, presumably low set of prices faced by the many farm families who have the option of growing their own food; official, subsidized prices set by the Polish government; and higher black market prices for the same or similar products. In fact, until 1988 the official prices of stables such as bread were set so low that very few farm families grew food for their own consumption, when queuing was not a problem.

Our estimates are based on a static consumption model and risk of bias due to the omission of dynamic factors such as habit persistence. We investigated this by estimating a dynamic version of our model that included lagged consumption. To obtain cross-sectional estimates of our dynamic model, we treated our data as though they came from three independent two-year panels (1984-85 through 1986-87 in the case of the PSID and 1987-88 through 1989-90 in the case of the Polish expenditure survey). We estimate this dynamic model as a system using SURE and both with and without instrumentation for log total income or expenditure. We found that the coefficient on the lagged dependent variables were significant in these dynamic models, but their inclusion changed the values of the short-run income elasticity very little.

An important result of this work is that pseudo panel estimates are often close to estimates based on genuine panel data. Large and similar apparent endogeneity biases were found in both countries. Cross-section estimations produce elasticities that are systematically higher for food at home and lower for food away. Moreover it seems that the aggregation lowers the endogeneity bias for food consumption. In further research these results should be verified on a complete set of expenditure data. Once corrected for the biases, food at home and food away income elasticities become very close to each other in the U.S. data, a result that seems reasonable to us and highlights possible errors that can arise from estimations using cross-sectional data.

## ACKNOWLEDGMENTS

We gratefully acknowledge helpful suggestions from participants at seminars at the University of Michigan, Northwestern University, Université de Paris I, Université de Genève, Erudite, CREST, Journées de Microéconomie Appliquée (1998), Congress of the European Economic Association, International Conference on Panel Data, as well as research support from Inra, Inrets and Credoc. The Polish data were available thanks to Professor Górecki, University of Warsaw, Department of Economics. We thank for a support from CREST, INSEE and MATISSE.

## APPENDIX A: HETEROSCEDASTICITY IN THE PSEUDO-PANEL MODEL

Write the pseudo-panel model presented in the text in a matrix form as:

$$y = X\beta + Z\alpha + \varepsilon$$

where $y$ is a NT column vector with N the number of cohorts and T the number of periods; $X$ is a matrix of explanatory variables and $\beta$ a vector of parameters; $Z$ is a (NT x N) matrix that contains cell dummies and $\alpha$ a vector of cohort effects; and $\varepsilon$ is a heteroscedastic residual. Call $D$ a diagonal (NT x NT) matrix such that $D = diag(\delta_{Ht})$ where $\delta_{Ht}$ is defined in the text and varies with cell size and the within-cell structure of relative expenditure. The model variance matrix is thus:

$$\Omega = \sigma_\alpha^2 TB + \sigma_\varepsilon^2 D$$

where $B$ is the between transformation matrix. The matrix $D$ in this expression is the source of cell and time-varying heteroscedasticity.

The GLS estimator of $\beta$ is

$$\hat{\beta}_{GLS} = (X'\Omega^{-1}X)^{-1}X'\Omega^{-1}y$$

In panel analysis, this can be expressed as a weighted sum of between and within estimators, which is called spectral decomposition. This holds if $\Omega^{-1}$ can be projected into the within and between dimensions, that is if matrices $\Omega_1$ and $\Omega_2$ exist such that

$$\Omega^{-1} = W\Omega_1 W + B\Omega_2 B$$

Gurgand, Gardes and Bolduc (1997) show that this can only be the case if matrix $B\Omega$ is symmetric, which implies that the weights in $D$ are time invariant. The intuition of this result is that if, in the process of scaling the variances, individual effects $\alpha$ receive differing weights for various observations of a given cohort, the within transformation no longer resolves individual heterogeneity, so that it does not strictly reflect the time-series variance of the model. Decomposing $\hat{\beta}_{GLS}$ into the cross-section and time-series dimensions based on the within and between estimates is thus no longer possible. In contrast, as long as cell size is constant over time, this argument does not hold and the decomposition obtains.

A corollary is that the efficient within estimator cannot be simply based on the original model weighted by heteroscedasticity time-varying factors, because this would create time-varying cohort effects. Gurgand, Gardes and Bolduc (1997) show that the within estimator that is both efficient and consistent when $\alpha$ and $X$ are correlated is

$$\hat{\beta}_W = (X'W\Delta WX)^{-1}X'W\Delta Wy$$

where $\Delta = D^{-1} - D^{-1}Z(Z'D^{-1}Z)^{-1}Z'D^{-1}$. Alternatively, when the number of cohorts is small, the least square dummy variable estimator can be used with weights directly proportional to $\delta_{Ht}$. The between estimator is straightforward because the time dimension is absent and is easily obtained by weighted least squares, with weights $(\sigma_\mu^2+\sigma_\varepsilon^2 w_c/T)^{-0.5}$ with $w_c = 1/T\ \Sigma_t\delta_{Ht}$.

## APPENDIX B: MEASURING THE SHADOW PRICES

Suppose that monetary price $p_m$ and a shadow price $\pi$ corresponding to non-monetary resources and to constraints faced by the households are combined together into a complete price. Expressed in logarithm form, we have: $p_c = p_m + \pi$.

Two estimations of the same equation : $x_{iht} = g(Z_{ht}) = Z_{ht}.\beta_i + u_{iht}$ (equation B.1) for good i (i = 1 to n), individual h (h = 1 to H) in period t (t = 1 to T) are made on cross-section and time-series over the same data-set. The residual is decomposed between $\alpha_{ih}$, the specific effect which contains all permanent components of the residual for individual h and good i, and the residual effect $\varepsilon_{iht}$: $u_{iht} = \alpha_{ih} + \varepsilon_{iht}$. The cross-section estimates can be biased by a correlation between some among the explanatory variables $Z_{ht}$ and this specific effect (see Mundlak, 1978). Such a correlation is due to latent permanent variables (such as an event during the infancy, characteristics of parents or permanent wealth) which are related both to the specific permanent effect and to the between transformation of the explanatory variables $Z_{ht}$. Note $\delta_i$ the correlation coefficient between the time average of the vector of the explanatory variables, $Z_{ht} = (Z^k_{ht})_{k=1 \text{ to } K1}$ , transformed by the Between matrix: $BZ_{ht} = \{(1/T)\ \Sigma_t Z^k_{ht}\}_{k=1 \text{ to } K1}$, and the specific effect: $\alpha_{ih} = BZ_{ht}.\delta_i + \eta_{ih}$, This coefficient adds to the parameter $\beta_i$ corresponding to the influence of Z in the between estimation : $Bx_{iht} = BZ_{ht}.(\beta_i + \delta_i) + \eta_{ih} + B\varepsilon_{iht}$, so that the between estimates are biased. Thus, the difference between the cross-section and the time-series estimates is equal to $\delta_i$.

Suppose that the shadow price depends on some among variables Z, for instance $Z^k_{ht}$.

The marginal propensity to consume with respect to $Z^k_{ht}$, when considering the effect of the shadow prices $\pi_{jht}$ on consumption, can be written:

$$dx_{iht}/dZ^k_{ht} = dg_i/dZ^k_{ht} + \Sigma_j\ (dg_i/d\pi_{jht}).(d\pi_{jht}/dZ^k_{ht}).$$

The component $\Sigma_j\ dg_i/d\pi_{jht}.d\pi_{jht}/dZ^k_{ht}$ of the marginal propensity of endogeneous variables can be used to identify the shadow price variation over $Z^k_{ht}$, $d\pi_{jht}/d\ Z^k_{ht}$, since it can be computed by resolving a system of n linear equations after having estimated the price marginal propensities $dg_i/d\pi_j = \gamma_{ij}$. In our estimations we consider only the *direct effect* through the price of good i : $\gamma_{ii}.d\pi_i/dZ^k_{ht}$ so that $d\pi_i/dZ^k_{ht} = [\beta_i^{(c.s.)} - \beta_i^{(t.s.)}]/\gamma_{ii}$ (equation B.2).

**Table 1: Income elasticities for food at home and away from home: Panel Study of Income Dynamics (1984-87)**

| | Between | Cross-Sections | Within | First-differences |
|---|---|---|---|---|
| Panel | | | | |
| Food at home | | | | |
| Not instrumented | 0.167 (.008) | 0.140 (.017) | -0.019 (.019) | -0.029 (.034) |
| Instrumented | | | | |
| - whole pop. | 0.010 (.027) | -0.030 (.054) | 0.363 (.082) | 0.473 (.158) |
| - robust est. | 0.187 (.025) | 0.134 (.050) | 0.383 (.077) | 0.447 (.150) |
| Food away | | | | |
| Not instrumented | 0.963 (.019) | 0.872 (.039) | -0.025 (.043) | 0.062 (.063) |
| Instrumented | | | | |
| - whole pop. | 1.046 (.043) | 0.966 (.094) | 0.375 (.166) | 0.431 (.260) |
| - robust est. | 1.050 (.044) | 0.960 (.095) | 0.387 (.166) | 0.447 (.258) |
| Surveys | 1984-85-86-87 | | | |
| N | 9720 | | | |
| Control variables | Age, equivalence scale and its square | | | |
| Pseudo-panel (Not instrumented) | | | | |
| Food at home (a) | 0.311 (.045) | 0.265 (.056) | 0.240 (.095) | 0.382 (.114) |
| (b) | 0.306 (.033) | | 0.460 (.062) | |
| Food away (a) | 1.387 (.068) | 1.265 (.083) | 0.800 (.150) | 0.847 (.172) |
| (b) | 1.384 (.060) | | 1.125 (.127) | |
| Surveys | 1984-85-86-87 | | | |
| N | 90 | | | |
| Control variables | Log of Age and its square, equivalence scale and its square | | | |

*Note:* The value in parenthesis are standard errors adjusted for the instrumentation of Total Expenditures by the usual method.

**Table 2: Total expenditure elasticities for food at home and away from home: Polish surveys (1987-90).**

| | Between | Cross-Sections | Within | First-differences |
|---|---|---|---|---|
| Panel | | | | |
| Food at home | | | | |
| Not instrumented | 0.579 (.004) | 0.536 (.009) | 0.466 (.005) | 0.451 (.010) |
| Instrumented | 0.494 (.010) | 0.567 (.023) | 0.755 (.013) | 0.788 (.028) |
| Food away | | | | |
| Not instrumented | 1.119 (.073) | 1.239 (.164) | 2.618 (.556) | 1.460 (.185) |
| Instrumented | 1.216 (.143) | 1.326 (.329) | 4.195 (1.080) | 1.315 (.393) |
| N | 14520 | 14520 | 14520 | 10890 |
| Control variables | Log of Age, proportion of children, Education level, Location, Log of relative price for all commodities, cross quarterly and yearly dummies | | | |
| Pseudo-panel (Not instrumented) | | | | |
| Food at home (a) | 0.583 (.011) | 0.572 (.017) | 0.549 (.020) | 0.864 (.033) |
| (b) | 0.591 (.010) | | 0.584 (.022) | |
| (c) | 0.591 (.011) | 0.581 (.018) | 0.526 (.020) | 0.568 (.033) |
| (d) | 0.589 (.013) | 0.581 (.018) | 0.965 (.023) | 0.915 (.032) |
| Food away (a) | 0.820 (.203) | 0.890 (.258) | - 0.218 (.318) | 0.696 (.331) |
| (b) | 0.609 (.208) | | -0.529 (.331) | |
| (c) | 0.608 (.213) | 0.240 (.270) | -0.072 (.322) | 0.333 (.508) |
| (d) | 1.149 (.212) | 0.367 (.268) | 0.624 (.199) | 0.965 (.315) |
| Surveys | 1987-88-89-90 | | | |
| N | 224 | | | |
| Control variables | Log of Age, proportion of children, Location, Log of relative prices for food, quarterly and yearly dummies | | | |

*Note:* The value in parenthesis are standard errors adjusted for the instrumentation of Total Expenditures by the usual method.

**Table 3: Hausman test for income parameters (food at home)**

| | panel | | Pseudo-panel |
|---|---|---|---|
| Instrument | without IV | with IV | Without IV |
| PSID | 107.7 | 46.8 | 0.5 |
| Poland | 336.6 | 292.4 | 2.2 |
| Poland (QAIDS) | 413.7 | 343.4 | - |

**Table 4: Income Elasticity of Food Shadow Prices**

| | | PSID (U.S.) | | Polish Panel | |
|---|---|---|---|---|---|
| Period | | 1984-87 | | 1987-90 | |
| N | | 2430 | | 3630 | |
| Prices | | No | | By social category | |
| Income Elasticity | | CS | TS | CS | TS |
| Food at Home | | 0.19 | 0.38 | 0.49 | 0.76 |
| Food Away | | 1.00 | 0.39 | 1.22 | 0.36 |
| Direct Price Elasticity (FH/FA) | | -0.19 | | -0.38/-0.18 | |
| Income Elasticity of the Shadow Price | (i) F.H. | 1.00 | | 0.71 | |
| | (ii) F.A. | -3.13 | | -4.78 | |

**Appendix Table 1: Means and standard deviations of variable used in the PSID analyses**

| | 1983 | 1984 | | 1985 | | 1986 | | 1987 | |
|---|---|---|---|---|---|---|---|---|---|
| | Level | Level | Dif. | Level | Dif. | Level | Dif. | Level | Dif. |
| Budget share for food at home | .147<br>(.103) | .144<br>(.098) | -.003<br>(.084) | .129<br>(.095) | -.015<br>(.086) | .137<br>(.100) | .008<br>(.082) | .134<br>(.096) | -.003<br>(.081) |
| % with at-home share = 0 | 0.0 | 0.0 | 53.2 | 0.0 | 74.0 | 0.0 | 41.5 | 0.0 | 51.3 |
| Budget share for food away from home | .033<br>(.040) | .034<br>(.038) | .001<br>(.034) | .031<br>(.038) | -.003<br>(.033) | .033<br>(.041) | .002<br>(.032) | .033<br>(.034) | .001<br>(.033) |
| % with away-from-home share =0 | 9.5 | 8.9 | 5.7 | 9.6 | 5.5 | 10.3 | 5.5 | 8.9 | 5.7 |
| ln household income | 9.9254<br>(.648) | 9.9985<br>(.657) | .0731<br>(.280) | 10.1714<br>(.716) | .1729<br>(.320) | 10.1238<br>(.686) | -.0475<br>(.308) | 10.1671<br>(.694) | .0432<br>(.299) |
| ln age Head | 3.7044<br>(.377) | 3.7306<br>(.368) | .0262<br>(.013) | 3.7573<br>(.359) | .0267<br>(.013) | 3.7801<br>(.351) | .0228<br>(.012) | 3.8044<br>(.343) | .0242<br>(.012) |
| ln family size (Oxford scale) | .6741<br>(.404) | .6837<br>(.401) | .0096<br>(.162) | .6896<br>(.405) | .0060<br>(.168) | .6894<br>(.409) | -.0002<br>(.159) | .6912<br>(.410) | .0018<br>(.171) |

*Note:* The value in parenthesis are standard errors.

**Appendix Table 2: Regression Coefficient and Standard Errors for Instrumental Variables Equation for Income Level for the PSID** (Dependent Variable: Disposable Family Income in logs in 1987, 1986, 1985)

| Independent Variable | Coefficient (Standard Error) | Independent Variable | Coefficient (Standard Error) |
|---|---|---|---|
| Quit $_{t}$ | -.049 (.014) | Birth | .001 (.015) |
| Quit $_{t-1}$ | -.036 (.014) | Age Head | .073 (.004) |
| Quit $_{t-2}$ | -.019 (.014) | Age Head squared | -.0007 (.00004) |
| Lay off $_{t}$ | -.066 (.021) | Wage growth*Quit $_{t}$ | -.063 (.033) |
| Lay off $_{t-1}$ | -.103 (.022) | Wage growth*Quit $_{t-1}$ | -.003 (.035) |
| Lay off $_{t-2}$ | -.053 (.020) | Wage growth*Quit $_{t-2}$ | -.005 (.032) |
| Promoted $_{t}$ | .025 (.022) | Wage growth*Lay off $_{t}$ | -.097 (.047) |
| Promoted $_{t-1}$ | .047 (.021) | Wage growth*Lay off $_{t-1}$ | -.093 (.051) |
| Promoted $_{t-2}$ | .017 (.021) | Wage growth*Lay off $_{t-2}$ | .005 (.042) |
| Unemp hrs $_{t}$ | -.271 (.044) | Wage growth*Promoted $_{t}$ | .043 (.074) |
| Unemp hrs $_{t-1}$ | .043 (.043) | Wage growth*Promoted $_{t-1}$ | -.149 (.075) |
| Unemp hrs $_{t-2}$ | .139 (.040) | Wage growth*Promoted $_{t-2}$ | -.035 (.069) |
| Hrs lost ill $_{t}$ | .266 (.044) | Region $_{1-2}$ | .037 (.021) |
| Hrs lost ill $_{t-1}$ | .387 (.047) | Region $_{>3}$ | .034 (.028) |
| Hrs lost ill $_{t-2}$ | .446 (.049) | City Size $_{1-2}$ | -.016 (.026) |
| Wage growth $_{t}$ | .050 (.024) | City Size $_{>3}$ | -.054 (.026) |
| Wage growth $_{t-1}$ | -.005 (.022) | Education Head | .138 (.006) |
| Wage growth $_{t-2}$ | -.005 (.016) | Wage | .012 (.003) |
| Divorce | .037 (.029) | Wage $_{t-1}$ | .022 (.003) |
| Marriage | .202 (.030) | | |

*Note:* The value in parenthesis are standard errors.

**Appendix Table 3: Means and standard deviations of variable used in the Polish panel analyses**

| | 1987 | | 1988 | | 1989 | | 1990 | |
|---|---|---|---|---|---|---|---|---|
| | Level | Dif. | Level | Dif. | Level | Dif. | Level | Dif. |
| Budget share for food at home | 0.508<br>(.14) | - | 0.484<br>(.15) | -0.024<br>(.14) | 0.486<br>(.18) | 0.003<br>(.17) | 0.554<br>(.15) | 0.068<br>(.17) |
| % with at-home share > 0 | 100 | - | 100 | 100 | 100 | 100 | 100 | 100 |
| Budget share for food away from home | 0.006<br>(.02) | - | 0.006<br>(.03) | -0.001<br>(.02) | 0.005<br>(.02) | -0.001<br>(.02) | 0.005<br>(.03) | .0002<br>(.03) |
| % with away-from-home share = 0 | 28.4 | - | 29.7 | - | 26.9 | - | 20.5 | - |
| ln household expenditure | 10.65<br>(.45) | - | 11.17<br>(.49) | 0.50<br>(.38) | 12.25<br>(.79) | -0.18<br>(.62) | 14.14<br>(.50) | -0.03<br>(.58) |
| ln head's age | 3.789<br>(.33) | - | 3.809<br>(.32) | 0.020<br>(.16) | 3.824<br>(.32) | 0.014<br>(.15) | 3.842<br>(.32) | 0.019<br>(.15) |
| ln family size | 1.140<br>(.59) | - | 1.121<br>(.60) | -0.019<br>(.24) | 1.095<br>(.61) | -0.026<br>(.21) | 1.081<br>(.61) | -0.014<br>(.22) |

*Note:* The value in parenthesis are standard errors.

**Appendix Table 4: Regression Coefficient and Standard Errors for Instrumental Variables Equation for Total Expenditure Level and Change for the Polish Expenditure Panel**

| Independent Variable | Coefficient (Standard Error) |
|---|---|
| log income | .374 (.071) |
| children % in family | -.270 (.024) |
| log age | 1.704 (.311) |
| log age squared | -.0299 (.042) |
| location (ref: countryside) | - - - - - - - - - |
| large city | .041 (.014) |
| average city | .028 (.014) |
| small city | .017 (.019) |
| social category (ref: wage earners) | - - - - - - - - - -- |
| wage earners-farmers | -.096 (.015) |
| Pensioners | -.047 (.017) |
| Farmers | -.196 (.018) |
| Education | -.028 (.003) |
| log income squared | .004 (.003) |

*Note:* The value in parenthesis are standard errors.